\documentclass[manuscript]{aastex}

\usepackage{spr-astr-addons}
\usepackage{color}

\begin{document}

\title{The moment of inertia of the neutron star PSR J0348+0432 and its proto neutron star}

\author{Xian-Feng Zhao$^{1,2}$}

\slugcomment{Not to appear in Nonlearned J., 45.}
\shorttitle{Short article title}
\shortauthors{Autors et al.}

\affil{$^{1}$ School of Sciences, Southwest Petroleum University, \\
{Chengdu, 610500, China}\\
$^{2}$ School of Electronic and Electrical Engineering,\\
{Chuzhou University, Chuzhou, 239000, China}\\
   e-mail:zhaopioneer.student@sina.com
   }

\begin{abstract}
The difference of the moment of inertia of the neutron star PSR J0348 + 0432 and that of its proto neutron star is studied in the framework of the relativistic mean field theory considering baryon octet. The temperature of the proto neutron star PSR J0348+0432 is chosen as T=5 MeV. The calculations show that the central baryon number density of the proto neutron star PSR J0348+0432 is in the range 0.623$\sim$0.813 fm$^{-3}$, decreased by 2$\sim$7\% compared to that of the neutron star PSR J0348+0432. The radius of the proto neutron star PSR J0348+0432 is in the range 13.101$\sim$12.419 km, increased by 1$\sim$2\% compared to that of the neutron star PSR J0348+0432. The moment of inertia of the proto neutron star PSR J0348+0432 is in the range 1.939$\times$10$^{45}$$\sim$1.638$\times$10$^{45}$ g.cm$^{2}$, increased by about 2$\sim$7\% compared to that of the neutron star PSR J0348+0432.
\end{abstract}


\section{Introduction}
Neutron star (NS) is a high-speed rotating object~\citep{Glendenning97}. To describe the rotational nature of NSs, we should know its moment of inertia, which can be calculated by the equations derived from the general theory of relativity~\citep{Hartle67,Hartle68}.

The equation of state (EoS) of the dense matter would influence the moment of inertia, conversely, the measurements of the moment of inertia (e.g. PSR J0737-3039A) can also constrain on the dense matter EoS~\citep{Bejger05}. A measurement of the moment of inertia of PSR J0737-3039A with 10\% error, without any other information from observations, will constrain the EoS over a range of densities to within
50\%$\sim$60\%. Tidal deformabilities between 0.6 and 6$\times$10$^{36}$ g.cm$^{2}$s$^{2}$ (to 95\% confidence) for M = 1.4 M$\odot$, and any measurement which constrains this range will provide an important
constraint on dense matter~\citep{Steiner15}. If the EoS is trusted up to the nuclear saturation density, a measurement of the moment of inertia will place absolute bounds on the radius of PSR J0737-3039A to within $\pm$1 km~\citep{Raithel16}.

In addition, the moment of inertia of the crust of a NS can explain the observed glitch activity of young pulsars arising from the exchange of angular momentum between the crust and the interior of the star~\citep{Atta15}.

Recent years, massive NSs have been discovered, such as NS PSR J1614-2230~\citep{Demorest10} and NS PSR J0348+0432~\citep{Antoniadis13}. A great number of theoretical calculations have been made for these massive NSs~\citep{zhaoprc12,zhaoprc15}.

But on the moment of inertia of the massive proto neutron star (PNS), which is very important for the understanding of the stellar evolution, there is little work to do. The reason may be that the life of the PNS is very short and to observe it will be difficult. For examle, for a PNS with mass $M$=1.4 M$_{\odot}$, its temperature $T$ will change from 30 MeV to 5 MeV within $t$=0$\sim$20 s~\citep{Burrows86}. For the massive PNSs, the cases should be similar.

In this paper, we study the difference of the moment of inertia $I$ of the NS PSR J0348+0432 and that of its PNS in the framework of the relativistic mean field (RMF) theory.

\section{The RMF theory and the parameters}
The Lagrangian density of hadron matter reads as follows~\citep{Glendenning97}
\begin{eqnarray}
\mathcal{L}&=&
\sum_{B}\overline{\Psi}_{B}(i\gamma_{\mu}\partial^{\mu}-{m}_{B}+g_{\sigma B}\sigma-g_{\omega B}\gamma_{\mu}\omega^{\mu}
\nonumber\\
&&-\frac{1}{2}g_{\rho B}\gamma_{\mu}\tau\cdot\rho^{\mu})\Psi_{B}+\frac{1}{2}\left(\partial_{\mu}\sigma\partial^{\mu}\sigma-m_{\sigma}^{2}\sigma^{2}\right)
\nonumber\\
&&-\frac{1}{4}\omega_{\mu \nu}\omega^{\mu \nu}+\frac{1}{2}m_{\omega}^{2}\omega_{\mu}\omega^{\mu}-\frac{1}{4}\rho_{\mu \nu}\cdot\rho^{\mu \nu}+\frac{1}{2}m_{\rho}^{2}\rho_{\mu}\cdot\rho^\mu
\nonumber\\
&&-\frac{1}{3}g_{2}\sigma^{3}-\frac{1}{4}g_{3}\sigma^{4}+\sum_{\lambda=e,\mu}\overline{\Psi}_{\lambda}\left(i\gamma_{\mu}\partial^{\mu}
-m_{\lambda}\right)\Psi_{\lambda}
.\
\end{eqnarray}
Here, the RMF theory is used~\citep{Glendenning97}. The energy density $\varepsilon$ and the pressure $p$ of a PNS are seen in Refs~\citep{Glendenning87} and those of a NS in Refs~\citep{Glendenning97}. The Tolman-Oppenheimer-Volkoff (TOV) equation and moment-of-inertia equation are used to obtain the mass, the radius and the moment of inertia of a NS/PNS~\citep{Glendenning97}.

\section{Parameters}
In this work, the nucleon coupling constant is chosen as the GL85 set~\citep{Glendenning85}: the saturation density $\rho_{0}$=0.145 fm$^{-3}$, binding energy B/A=15.95 MeV, a compression modulus $K=285$ MeV, charge symmetry coefficient $a_{sym}$=36.8 MeV and the effective mass $m^{*}/m$=0.77.

We define the ratios of hyperon coupling constants to nucleon coupling constants as follows: $x_{\sigma h}=\frac{g_{\sigma h}}{g_{\sigma}}$, $x_{\omega h}=\frac{g_{\omega h}}{g_{\omega}}$, $x_{\rho h}=\frac{g_{\rho h}}{g_{\rho}}
$, with $h$ denoting hyperons $\Lambda, \Sigma$ and $\Xi$.

We select $x_{\rho \Lambda}=0, x_{\rho \Sigma}=2, x_{\rho \Xi}=1$ by SU(6) symmetry~\citep{Schaff96}. According to the experimental data, the hyperon well depth is $U_{\Lambda}^{(N)}=-30$ MeV~\citep{Batt97}, $ U_{\Sigma}^{(N)}=10\sim40$ MeV~\citep{{Kohno06},{Harada05},{Harada06},{Fried07}} and $U_{\Xi}^{(N)}=-18$ MeV~\citep{Schaff00}, respectively. Then, $U_{\Lambda}^{(N)}=-30$ MeV, $ U_{\Sigma}^{(N)}$=+30 MeV and $U_{\Xi}^{(N)}=-18$ MeV are chosen in this work.

The ratio of hyperon coupling constant to nucleon coupling constant is in the range of $\sim$ 1/3 to 1~\citep{Glen91}. Therefore, we choose $x_{\sigma \Lambda}$=0.4, 0.5, 0.6, 0.7, 0.8, 0.9 at first in this work. The hyperon coupling constants $x_{\omega \Lambda}$ can be obtained by considering the restriction of the hyperon well depth~\citep{Glendenning97}

\begin{eqnarray}
U_{h}^{(N)}=m_{n}\left(\frac{m_{n}^{*}}{m_{n}}-1\right)x_{\sigma h}+\left(\frac{g_{\omega}}{m_{\omega}}\right)^{2}\rho_{0}x_{\omega h}
.\
\end{eqnarray}

Thus we obtain parameter ($x_{\sigma \Lambda}$, $x_{\omega \Lambda}$)=(0.4, 0.3681; 0.5, 0.5090; 0.6, 0.6500; 0.7, 0.7910; 0.8, 0.9320) (named as parameter $\alpha$). Similarly, we can obtain  ($x_{\sigma \Sigma}$, $x_{\omega \Sigma}$)=(0.4, 0.7597; 0.5, 0.9007) (named as parameter $\beta$) and ($x_{\sigma \Xi}$, $x_{\omega \Xi}$)=(0.4, 0.4464; 0.5, 0.5874; 0.6, 0.7284; 0.7, 0.8693) (named as parameter $\gamma$).

We can get 40 sets of parameters by taking each from parameters $\alpha$, $\beta$ and $\gamma$ (named as No.01, No.02, ... , No.40, respectively).

In order to study the nuance in nature between the PNS PSR J0348+0432 and the NS PSR J0348+0432, we select the temperature $T$=5 MeV in this work.

\section{The mass and radius of the NS/PNS PSR J0348+0432}
\begin{figure}[!htp]
\begin{center}
\includegraphics[width=3.3in]{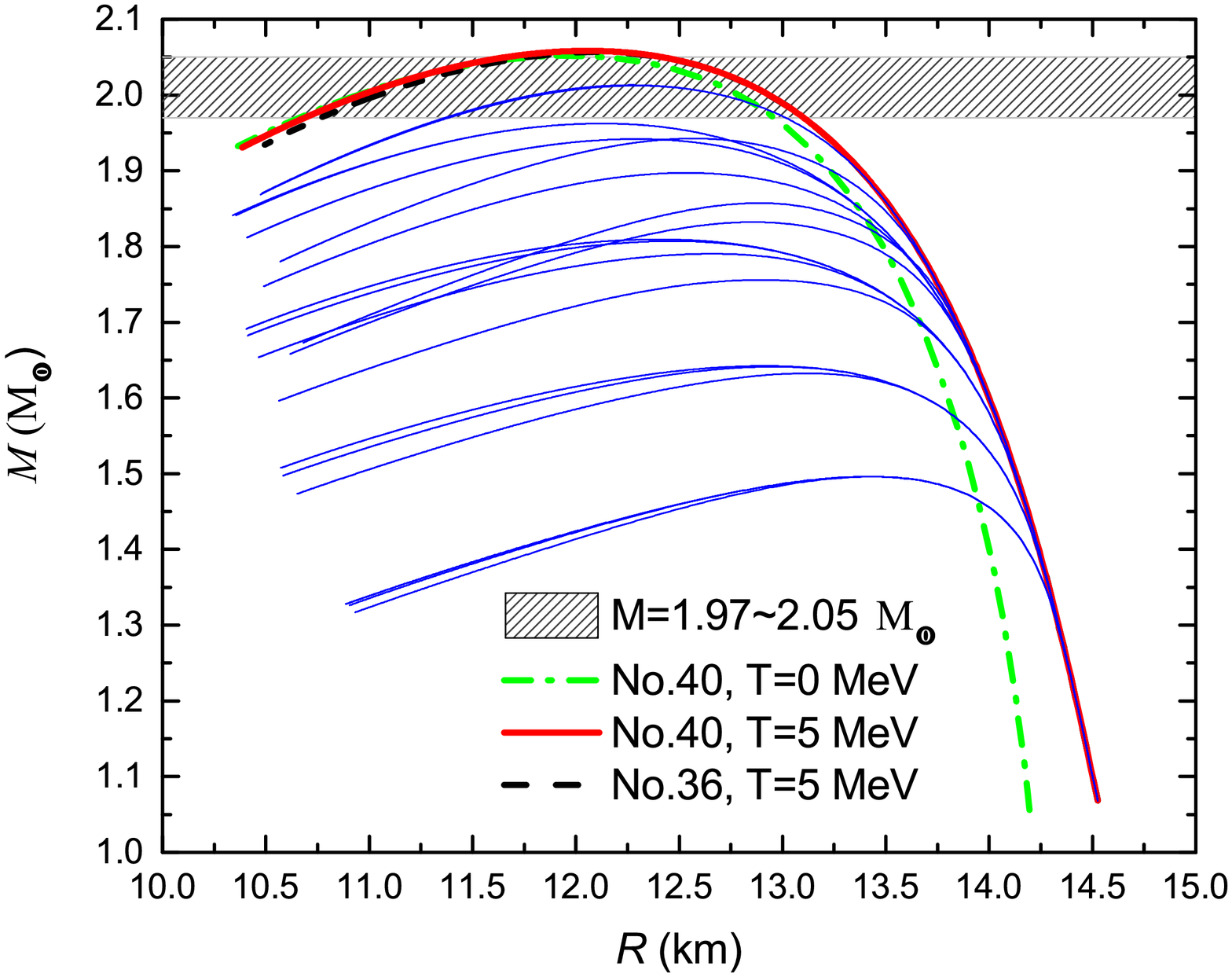}
\caption{The mass $M$ of a NS/PNS as a function of the radius $R$.}
\label{fig1}
\end{center}
\end{figure}

For parameters No.01$\sim$No.40, we calculate the mass of the PNS PSR J0348+0432 (see Fig.~\ref{fig1}). We see that parameters No.36 ($x_{\sigma \Lambda}$=0.8, $x_{\omega \Lambda}$=0.9320; $x_{\sigma \Sigma}$=0.4, $x_{\omega \Sigma}$=0.7597; $x_{\sigma \Xi}$=0.7, $x_{\omega \Xi}$=0.8693) and No.40 ($x_{\sigma \Lambda}$=0.8, $x_{\omega \Lambda}$=0.9320; $x_{\sigma \Sigma}$=0.5, $x_{\omega \Sigma}$=0.9007; $x_{\sigma \Xi}$=0.7, $x_{\omega \Xi}$=0.8693) can give the mass of the PNS PSR J0348+0432. For parameter No.40, the maximum mass of the NS PSR J0348+0432 is $M$=2.052 M$_{\odot}$. So, we can use parameters No.40 to describe the PNS PSR J0348+0432.

\begin{figure}[!htp]
\begin{center}
\includegraphics[width=3.3in]{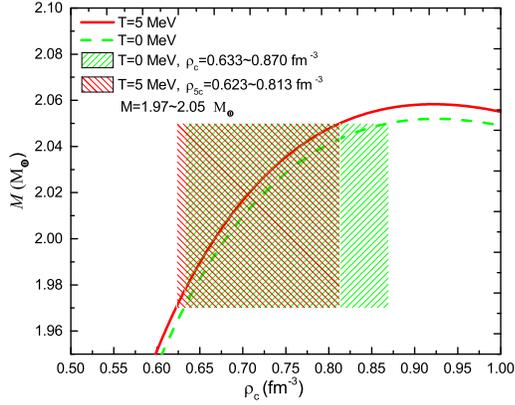}
\caption{The mass $M$ of the NS/PNS as a function of the central baryon number density $\rho_{c}$.}
\label{fig2}
\end{center}
\end{figure}

The mass $M$ of the NS/PNS J0348+0432 as a function of the central baryon number density $\rho_{c}$ is shown in Fig~\ref{fig2}. We see that the baryon number density of the PNS J0348+0432 is in the range $\rho_{5c}$=0.623$\sim$0.813 fm$^{-3}$, which is smaller than that of the NS J0348+0432 ($\rho_{c}$=0.633$\sim$0.870 fm$^{-3}$). Temperature effect makes the baryon number density of the PNS J0348+0432 increased by about 2$\sim$7\% compared to that of the NS J0348+0432. These also can be seen in Table~\ref{tab1}.

\begin{table}[t]
\centering
\caption{The radius $R$ and the moment of inertia $I$ of the NS/PNS PSR J0348+0432 calculated in this work. Here, the mass of the NS/PNS is in the range $M$=1.97$\sim$2.05 M$_{\odot}$.}
\label{tab1}
\begin{tabular}{llll}
\hline\noalign{\smallskip}
$T$   &$\rho_{c}$                          &$R$               &$I$   \\
MeV   &fm$^{-3}$       &km                &$\times$10$^{45}$ g.cm$^{2}$\\
\hline
0(NS)   &0.633$\sim$0.870               &12.957$\sim$12.143&1.906$\sim$1.537\\
5(PNS)  &0.623$\sim$0.813           &13.101$\sim$12.419&1.939$\sim$1.638\\
\noalign{\smallskip}\hline
\end{tabular}
\vspace*{0.5cm}  
\end{table}

\begin{figure}[!htp]
\begin{center}
\includegraphics[width=3.3in]{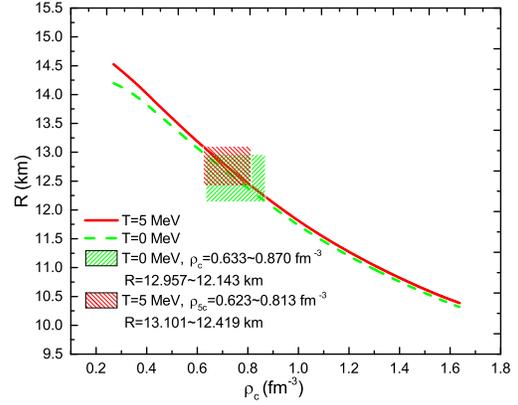}
\caption{The radius $R$ of the NS/PNS J0348+0432 as a function of the central baryon number density $\rho_{c}$}
\label{fig3}
\end{center}
\end{figure}

The radius $R$ of the NS/PNS J0348+0432 as a function of the central baryon number density $\rho_{c}$ is shown in Fig~\ref{fig3}, from which and Table~\ref{tab1} we see the radius of the PNS PSR J0348+0432 is in the range $R_{5}$=13.101$\sim$12.419 km, which is larger than that of the NS PSR J0348+0432 ($R$=12.957$\sim$12.143 km). The radius of the PNS PSR J0348+0432 increases by about 1$\sim$2\% compared to that of the NS PSR J0348+0432.

\section{The moment of inertia of the NS/PNS PSR J0348+0432}
\begin{figure}[!htp]
\begin{center}
\includegraphics[width=3.3in]{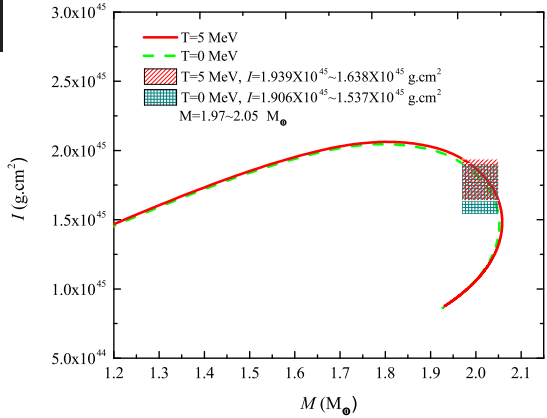}
\caption{The moment of inertia of the NS/PNS PSR J0348+0432 as a function of the mass.}
\label{fig4}
\end{center}
\end{figure}

The moment of inertia of the NS/PNS PSR J0348+0432 as a function of the mass is shown in Fig~\ref{fig4}. The upper half curve represents stable NS/PNS and the lower half curve represents unstable ones. We see that the moment of inertia of the PNS PSR J0348+0432 is larger than that of the NS PSR J0348+0432 corresponding to a same baryon number density $\rho_{c}$. If $I_{5}$ is defined as the moment of inertia of the PNS PSR J0348+0432 with the temperature of $T$=5 MeV, we also see that the moment of inertia of the PNS is in the range $I_{5}$=1.939$\times$10$^{45}$$\sim$1.638$\times$10$^{45}$ g.cm$^{45}$ while that of the NS PSR J0348+0432 is in the range $I$=1.906$\times$10$^{45}$$\sim$1.537$\times$10$^{45}$ g.cm$^{45}$. Temperature effect makes the moment of inertia of the PNS PSR J0348+0432 increased by about 2$\sim$7\% compared to that of the NS PSR J0348+0432.

\begin{figure}[!htp]
\begin{center}
\includegraphics[width=3.3in]{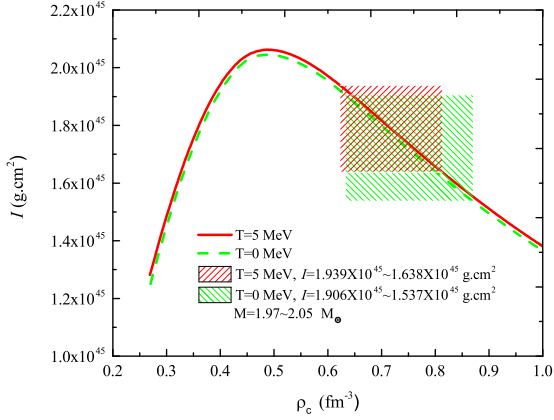}
\caption{The moment of inertia of the NS/PNS PSR J0348+0432 as a function of the central baryon number density $\rho_{c}$.}
\label{fig5}
\end{center}
\end{figure}

In order to study the relationship between the moment of inertia $I$ and the central baryon number density $\rho_{c}$, the moment of inertia of the NS/PNS PSR J0348+0432 as a function of the central baryon number density $\rho_{c}$ is shown in Fig.~\ref{fig5}. The right half curve represents stable NS/PNS and the left half curve represents unstable ones.

We see the moment of inertia of the NS/PNS PSR J0348+0432 decreases with the increase of the central baryon number density $\rho_{c}$. We also see that the moment of inertia of the PNS PSR J0348+0432 is larger than that of the NS PSR J0348+0432 corresponding to a same baryon number density $\rho_{c}$.

\section{Conclusions}
In this paper, the difference between the moment of inertia of the NS PSR J0348+0432 and its PNS is studied with the RMF theory.

The calculations show that the radius of the PNS PSR J0348+0432 is in the range $R_{5}$=13.101$\sim$12.419 km while that of the NS PSR J0348+0432 is in the range $R$=12.957$\sim$12.143 km, which is consistent with the results of Steiner et al~\citep{Steiner15}. The radius of the PNS PSR J0348+0432 increases by about 1$\sim$2\% compared to that of the NS PSR J0348+0432.

We also see that the moment of inertia of the PNS PSR J0348+0432 is in the range $I_{5}$=1.939$\times$10$^{45}$ $\sim$ 1.638$\times$10$^{45}$ g.cm$^{45}$ while that of the NS PSR J0348+0432 is in the range $I$=1.906$\times$10$^{45}$$\sim$1.537$\times$10$^{45}$ g.cm$^{45}$. The moment of inertia of the PNS PSR J0348+0432 increases by about 2$\sim$7\% compared to that of the NS PSR J0348+0432.

\section*{Acknowledgements}
We are thankful to Shan-Gui Zhou for fruitful discussions during my visit to the Institute of Theoretical Physics, Chinese Academy of Sciences.
This work was supported by the Natural Science Foundation of China (Grant No. 11447003) and the Scientific Research Foundation of the Higher Education Institutions of Anhui Province, China (Grant No. KJ2014A182).

\clearpage

\end{document}